\documentclass[a4paper,11pt]{article}
\usepackage{pos}
\usepackage{bm}
\usepackage{color}

\title{A proposal for removing $\pi N$-state contamination from the nucleon induced pseudoscalar form factor in lattice QCD}
\ShortTitle{A proposal for removing $\pi N$-state contamination from the nucleon induced pseudoscalar....}

\author*[a]{Shoichi Sasaki}
\author[b]{Yasumichi Aoki}
\author[c]{Ken-Ichi Ishikawa}
\author[d]{Yoshinobu Kuramashi}
\author[e]{Kohei Sato}
\author[d]{Eigo Shintani}
\author[f]{Ryutaro Tsuji}
\author[g]{Hiromasa Watanabe}
\author[h,d]{Takeshi Yamazaki}
\affiliation[]{\normalsize{\bf \sffamily \hspace{50mm}(PACS Collaboration)}}

\affiliation[a]{Department of Physics, Tohoku University, 980-8578, Sendai, Japan}

\affiliation[b]{RIKEN Center for Computational Science, 650-0047, Kobe, Japan}

\affiliation[c]{
Core of Research for the Energetic Universe,
Graduate School of Advanced Science and Engineering, Hiroshima University 739-8526, Higashi-Hiroshima, Japan}

\affiliation[d]{
Center for Computational Sciences, University of Tsukuba, 305-8577, Tsukuba, Japan}

\affiliation[e]{
Degree Programs in Pure and Applied Sciences, Graduate School of Science and Technology, University of Tsukuba, Ibaraki 305-8571, Japan
}

\affiliation[f]{High Energy Accelerator Research Organization (KEK),
  305-0801, Tsukuba, Japan}

\affiliation[g]{
Yukawa Institute for Theoretical Physics, Kyoto University, Kyoto 606-8502, Japan
}

\affiliation[h]{
Institute of pure and Applied Sciences, University of Tsukuba, 305-8571, Tsukuba, Japan
}

\emailAdd{ssasaki@nucl.phys.tohoku.ac.jp}

\abstract{
In the PACS10 project, the PACS collaboration has generated three sets of the PACS10 gauge configurations at the physical point with lattice volume larger than $(10\;{\rm fm})^4$ and three different lattice spacings. The isovector nucleon form factors had been already calculated by using two sets of the PACS10 gauge configurations. In our strategy, the smearing parameters of the nucleon interpolation operator were highly optimized to eliminate as much as possible the contribution of excited states in the nucleon two-point function. This strategy was quite successful in calculations of the electric ($G_E$), magnetic ($G_M$) and axial-vector ($F_A$) form factors, while the induced pseudoscalar ($F_P$) and pseudoscalar ($G_P$) form factors remained strongly affected by residual contamination of 
$\pi N$-state contribution. In this work, we propose a simple method to remove the $\pi N$-state contamination from the $F_P$ form factor, and then evaluate the induced pseudoscalar charge $g_P^\ast$ and the pion-nucleon coupling $g_{\pi NN}$ from existing data in a new analysis. Applying this method to the $G_P$ form factor is also considered with a help of the axial Ward-Takahashi identity.
}

\FullConference{%
  The 41st International Symposium on Lattice Field Theory (Lattice 2024)\\
  July 28th - August 3rd, 2024\\
  The University of Liverpool, United Kingdom
}

\begin{document}
\maketitle

\section{Introduction}

The axial structure of the nucleon is highly connected with the physics
of chiral symmetry and its spontaneous breaking, which ensures the presence of pseudo Nambu-Goldstone particles such as the pion. This is
empirically known as the partially conserved axial-vector current (PCAC) hypothesis, where the divergence of the axial-vector current
is proportional to the pion field. Applying this idea to the axial-vector matrix element of the nucleon given by
%
%
\begin{align}
\label{eq:NME_AV}
\langle N(p^\prime)|A_\alpha(x)|N(p)\rangle = \overline{u}_N(p^\prime)
\left[
\gamma_\alpha\gamma_5 F_A(q^2) +i q_\alpha \gamma_5 F_P(q^2)
\right]u_N(p)e^{iq\cdot x}
\end{align}
with $q=p -p^\prime$, a specific relation, known as the Goldberger-Treiman (GT) relation~\cite{Goldberger:1958vp}, is
derived between the axial-vector coupling defined by the axial-vector ($F_A$) form factor at $q^2=0$ and the residue of the pion-pole structure in the induced pseudo-scalar ($F_P$) form factor. 
Instead of PCAC, the axial Ward-Takahashi identity, $\partial_{\alpha} A_{\alpha}(x) = 2m P(x)$, leads to the generalized GT relation~\cite{{Weisberger:1966ip},{Sasaki:2007gw}}: 
%
%
\begin{align}
\label{eq:GGT}
2M_N F_A(q^2)-q^2 F_P(q^2)= 2m G_P(q^2),
\end{align}
which is satisfied among the three nucleon form factors including the pseudoscalar ($G_P$) form factor defined in the pseudoscalar matrix element of the nucleon
as
%
%
\begin{align}
\label{eq:NME_PS}
\langle N(p^\prime)|P(x)|N(p)\rangle = \overline{u}_N(p^\prime)
\left[
\gamma_5 G_P(q^2)
\right]u_N(p)e^{iq\cdot x}.
\end{align}
In addition, the following pion-pole dominance (PPD) ans\"atz~\cite{Nambu:1960xd}
for $F_P(q^2)$ and $G_P(q^2)$ at low $q^2$, 
%
%
\begin{align}
\label{eq:PPD}
F_P^{\mathrm{PPD}}(q^2)=\frac{2M_NF_A(q^2)}{q^2+m_\pi^2}\;\;\mbox{and}\;\;2m G_P^{\mathrm{PPD}}(q^2)=2M_NF_A(q^2)\frac{m_\pi^2}{q^2+m_\pi^2}, 
\end{align}
satisfies the generalized GT relation~(\ref{eq:GGT}).

Although the axial-structure of the nucleon has been studied extensively in lattice QCD at the physical point, no results have been obtained that satisfy the generalized GT relation well or give better accuracy than the PPD model. Indeed, in our previous works~\cite{{Shintani:2018ozy},{Tsuji:2023llh}}, both the $F_P$ and $G_P$ form factors are significantly underestimated in the low-$q^2$ region compared to the PPD model. This is simply 
due to strong $\pi N$ excited-state contamination. In this work, we propose a simple method to remove the $\pi N$-state contamination from the $F_P$ and $G_P$ form factors, and then evaluate the induced pseudoscalar charge $g_P^\ast=m_\mu F_P(0.88m_\mu^2)$ and the pion-nucleon coupling $g_{\pi NN}=\lim_{q^2\rightarrow \infty}(q^2+m_\pi^2)\frac{F_P(q^2)}{2F_\pi}$ from existing data in a new analysis.

\section{Standard method}
\label{sec:Std_method}

The nucleon two-point (2pt) function from the source-time position (denoted $t_{\mathrm{src}}$) to the sink-time position (denoted $t_{\mathrm{sink}}$) is defined as
%
%
\begin{align}
\label{eq:2ptC}
C_N(t_{\mathrm{src}}-t_{\mathrm{sink}};\bm{p})=\frac{1}{4}{\rm Tr}\left\{
{\cal P}_{+}\langle N(t_{\mathrm{sink}};\bm{p})
\overline{N}(t_{\mathrm{src}};-\bm{p})\rangle\right\}
\;\;\mbox{with}\;\;{\cal P}_{+}=\frac{1+\gamma_4}{2},
\end{align}
where the nucleon operator $N(t;\bm{q})$ carrying a three-dimensional momentum $\bm{p}$. 

The nucleon form factors are extracted from the
nucleon three-point (3pt) function consisting of the nucleon source and sink operators
with a given local current ($J$) defined as
%
%
\begin{align}
\label{eq:3ptC}
&C_{J}^{5z}(t; \bm{p}^{\prime}, \bm{p})=\frac{1}{4}{\rm Tr}\left\{
{\cal P}^{5z}\langle N(t_{\mathrm{sink}};\bm{p}^\prime)J(t;\bm{q})
\overline{N}(t_{\mathrm{src}};-\bm{p})
\right\},
\end{align}
where the projection operator ${\cal P}^{5z}={\cal P}_{+}\gamma_5\gamma_3$ is chosen for $J=A_{\alpha}$ or $P$.
We then calculate the following ratio constructed from an appropriate combination of the 2pt and 3pt functions~\cite{{Hagler:2003jd},{Gockeler:2003ay}} with a fixed source-sink separation ($t_{\mathrm{sep}}\equiv t_{\mathrm{sink}}-t_{\mathrm{src}}$):
%
%
\begin{align}
\mathcal{R}_{J}^{5z}(t; \bm{p}^{\prime}, \bm{p}) 
=\frac{C_{J}^{5z}(t; \bm{p}^{\prime}, \bm{p})}{C_{N}(t_{\mathrm{sink}}-t_{\mathrm{src}}; \bm{p}^{\prime})}\sqrt{\frac{C_{N}(t_{\mathrm{sink}}-t; \bm{p}) C_{N}(t-t_{\mathrm{src}}; \bm{p}^{\prime}) C_{N}(t_{\mathrm{sink}}-t_{\mathrm{src}}; \bm{p}^{\prime})}{C_{N}(t_{\mathrm{sink}}-t; \bm{p}^{\prime}) C_{N}(t-t_{\mathrm{src}}; \bm{p}) C_{N}(t_{\mathrm{sink}}-t_{\mathrm{src}}; \bm{p})}}.
\label{eq:ratio_3pt_2pt}
\end{align}
Since all $t$-dependence due to the contribution of the nucleon ground state can be eliminated in the ratio~(\ref{eq:ratio_3pt_2pt}), the target quantity can be read off from an asymptotic plateau of the ratio $\mathcal{R}_{J}^{5z}(t;\bm{p}^{\prime}, \bm{p})$,
being independent of the choice of $t_{\mathrm{sep}}$, if the condition $t_{\mathrm{sep}}/a \gg
(t-t_{\mathrm{src}})/a \gg 1$ is satisfied. In this study, we consider only the rest frame of the final state with $\bm{p}^\prime=\bm{0}$,
which leads to the condition of 
$\bm{q}=\bm{p}-\bm{p}^{\prime}=\bm{p}$. 
Therefore, 
the squared four-momentum transfer is given by $q^2=2M_N(E_N(\bm{q})-M_N)$ where $M_N$ and $E_N(\bm{q})$ represent the nucleon mass and energy with the momentum $\bm{q}$. 
In this kinematics, we use a simpler notation like 
$\mathcal{R}_{J}^{5z}(t; \bm{q})$ and $C_{J}^{5z}(t; \bm{q})$.

The ratio $\mathcal{R}_{J}^{5z}(t; \bm{q})$ 
gives the following asymptotic values including the respective form factors in the asymptotic region~\cite{Sasaki:2007gw}:
%
%
\begin{align}
    \label{eq:fa_def}
    \mathcal{R}^{5z}_{A_i}(t;\bm{q})
    &=
    K^{-1}
    \left[
        (E_N(\bm{q})+M_N)\widetilde{F}_A(q^2)\delta_{i3}-q_iq_3\widetilde{F}_P(q^2) 
    \right] + \cdot\cdot\cdot,\\
    \label{eq:fa4_def}
    \mathcal{R}^{5z}_{A_4}(t;\bm{q})
    &=iq_3 K^{-1}
    \left[
        \widetilde{F}_A(q^2)-(E_N(\bm{q})-M_N)\widetilde{F}_P(q^2)
    \right]+ \cdot\cdot\cdot, \\
    \label{eq:gp_def}
    \mathcal{R}_{P}^{5z}(t; \boldsymbol{q}) 
    & =iq_3 K^{-1}
    \widetilde{G}_{P}\left(q^{2}\right)+ \cdot\cdot\cdot,
\end{align}
with $K=\sqrt{2E_N(\bm{q})(E_N(\bm{q})+M_N)}$. The ellipsis denotes excited-state contributions, 
which are supposed to be ignored in the case of $t_{\mathrm{sep}}/a\gg (t-t_{\mathrm{src}})/a \gg 1$.
Three target quantities: $\widetilde{F}_A(q^2)$,  $\widetilde{F}_P(q^2)$ and $\widetilde{G}_P(q^2)$ ~\footnote{
Hereafter, the form factors with and without tilde denote the bare and renormalized ones, then
{\it e.g.} $F_A=Z_A \widetilde{F}_A$. 
} can be read off from an asymptotic plateau
of the ratio $\mathcal{R}_{J}^{5z}(t; \bm{q})$,
being independent of the choice of $t_{\mathrm{sep}}$. This approach is hereafter referred to as the 
standard method.

\section{Simple subtraction method}
\label{sec:New_method}

%
%
\begin{figure*}[ht]
\centering
\includegraphics[width=0.3\textwidth,bb=0 0 485 392,clip]{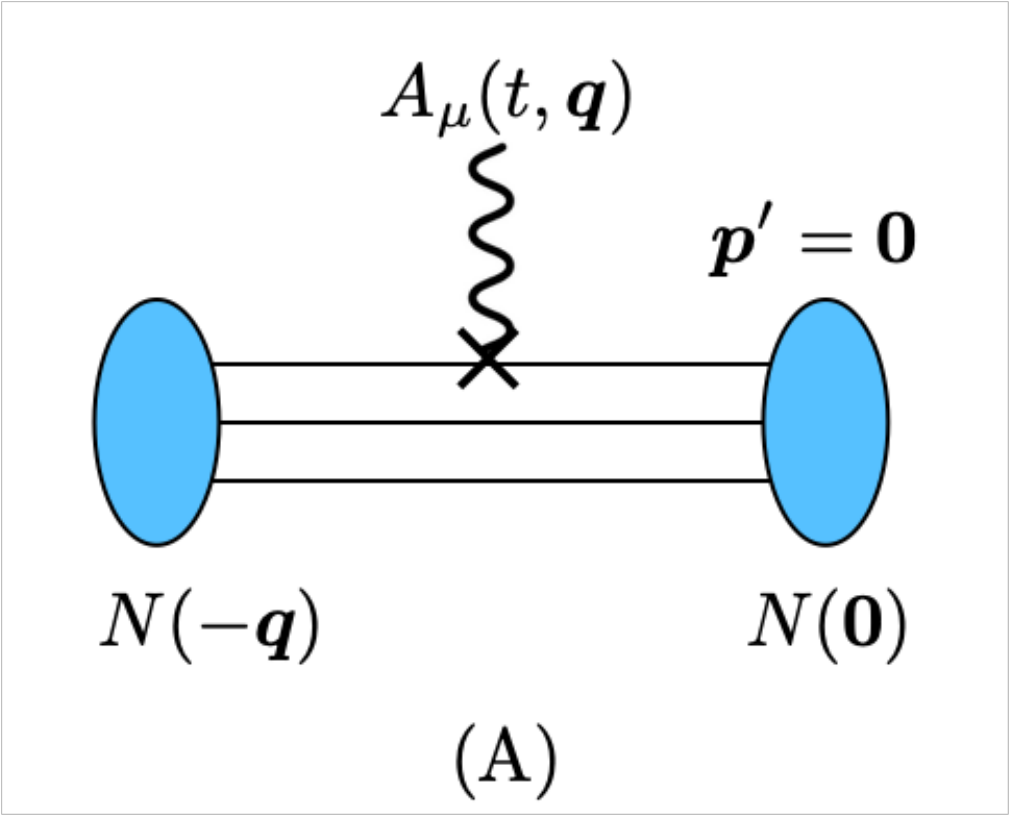}
\includegraphics[width=0.3\textwidth,bb=0 0 485 392,clip]{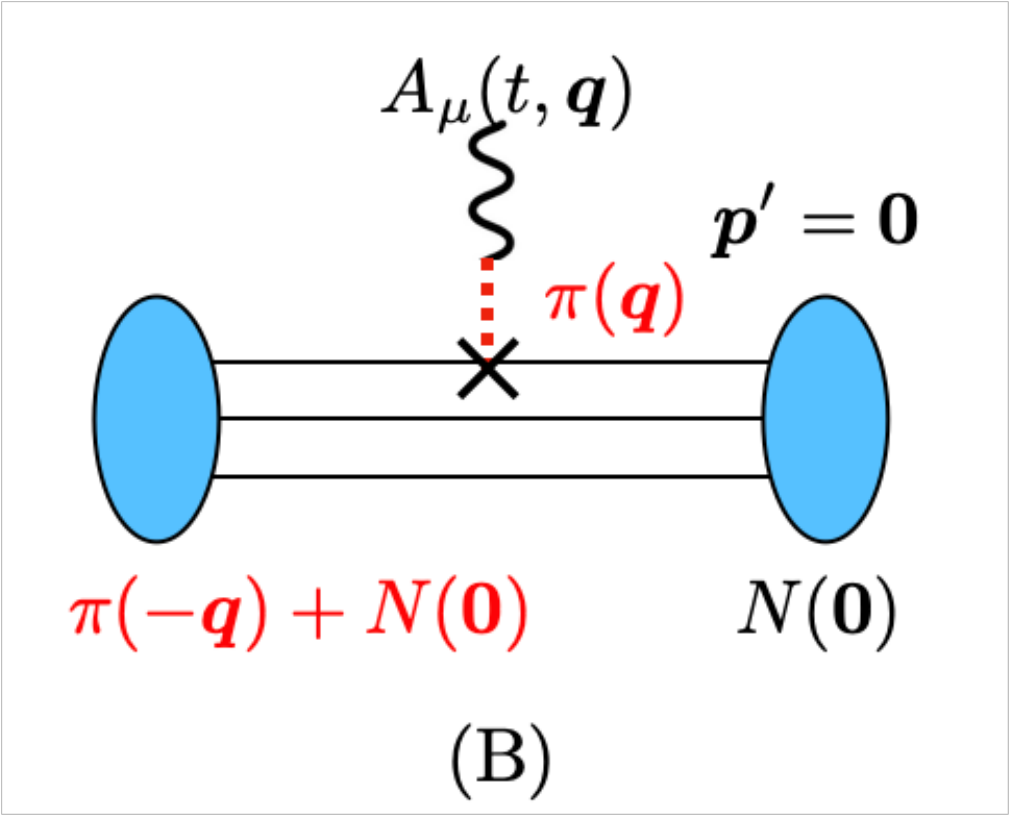}
\includegraphics[width=0.3\textwidth,bb=0 0 485 392,clip]{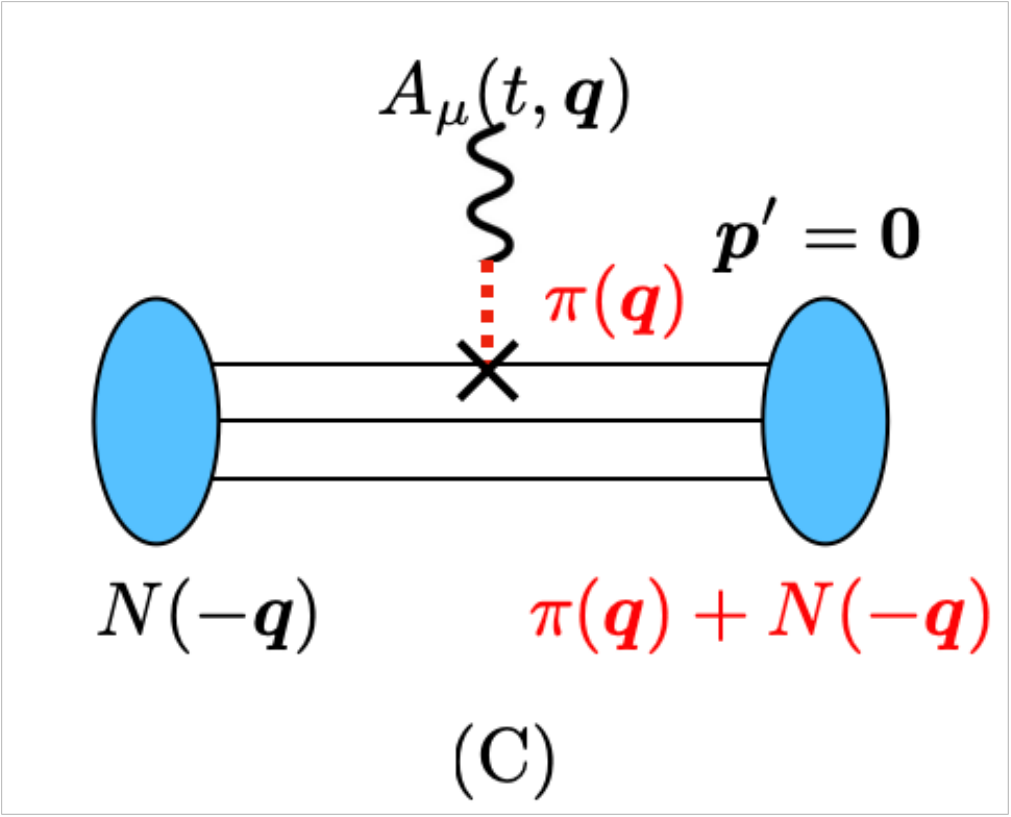}
\caption{Schematic view of the ground-state contribution \rm{(A)}
and two types of the leading $\pi N$ contributions \rm{(B)} and \rm{(C)} for the axial-vector matrix element.} \label{fig:quarkline_diag}
\end{figure*}

In our previous works~\cite{{Ishikawa:2018rew},{Shintani:2018ozy},{Ishikawa:2021eut},{Tsuji:2023llh}}, the 3pt functions involving the $A_4$ current are
not taken into account for the calculation of the $F_A(q^2)$ and $F_P(q^2)$
form factors.
This is simply because, to the best of our knowledge, the $A_4$ correlator was found to be statistically very noisy in Ref.~\cite{Ishikawa:2018rew}, where the time-reversal averaging was performed using both forward and backward propagation in time for all 3pt functions.
However, as pointed out for the first time in Ref.~\cite{Bali:2018qus}, 
the ratio correlator ${\cal R}^{5z}_{A_4}(t,\bm{q})$ does not show a plateau, but rather a peculiar behavior that depends linearly on the current insertion time $t$ with a steep negative slope under their kinematic setup.
When no time-reversal averaging is applied in our data, an almost linear $t$-dependence is indeed confirmed, giving the same slope, although the direction is reversed according to the respective kinematics.

As discussed in Ref.~\cite{Bar:2018xyi}, such peculiar time dependence is understood as the leading contribution from the $\pi N$ state in ${\cal R}^{5z}_{A_4}(t,\bm{q})$, arising in the tree diagram of the baryon ChPT. Importantly, the momentum $\bm{q}$ injected by the axial-vector current is entirely inherited by the pion state, since the pion in such $\pi N$ state remains in the on-mass shell. The kinematics of the leading $\pi N$ contribution is therefore restricted to two special cases as depicted in Fig.~\ref{fig:quarkline_diag} (B) and (C)~\cite{{Meyer:2018twz},{RQCD:2019jai}}.

In our previous works~\cite{{Ishikawa:2018rew},{Shintani:2018ozy},{Ishikawa:2021eut},{Tsuji:2023llh}}, 
the $F_P$ form factor obtained from $\tilde{\cal R}^{5z}_{A_i}(t,{\bm q})$ 
was indeed significantly affected by the excited-state contamination, though no such effect was observed for 
the $F_A$ form factor~\footnote{
Recent studies solving the generalized eigenvalue problem including $\pi N$ operators also show that the $\pi N$ contributions are strong in $\widetilde{F}_P(q^2)$ and $\widetilde{G}_P(q^2)$, 
but not in $\widetilde{F}_A(q^2)$~\cite{{Barca:2022uhi},{Alexandrou:2024tin}}.
}. Therefore, we assume that the contributions from the leading $\pi N$ state
for ${\cal R}^{5z}_{A_\alpha}(t,{\bm q})$ can be described as follows
%
%
\begin{align}
\tilde{\cal R}^{5z}_{A_i}(t,{\bm q})
&\equiv {\cal R}^{5z}_{A_i}(t,{\bm q})-\delta_{i3}{\cal R}^{5z}_{A_3}(t,{\bm q}_0)
= - q_3q_iK^{-1}\left[\widetilde{F}_P(q^2)-\Delta_{+}(t, t_{\mathrm{sep}}; {\bm q}) \right], \\
{\cal R}^{5z}_{A_4}(t,{\bm q})
&=iq_3K^{-1}\left[
\left(\widetilde{F}_A(q^2)-(E_N({\bm q})-M_N)\widetilde{F}_P(q^2)\right)+E_\pi({\bm q}) \Delta_{-}(t, t_{\mathrm{sep}}; {\bm q}) \right]
\end{align}
with $\bm{q}_0=(q_1, q_2, 0)$ satisfying $|\bm{q}_0|=|\bm{q}|$. The functions $\Delta_{\pm}(t, t_{\mathrm{sep}}; {\bm q})$ encode the leading $\pi N$ contributions, which
provide the residual $t$-dependence with a given $t_{\mathrm{sep}}$. 

For the case when the current operator carries the momentum ${\bm q}$, 
the $\pi N$ contribution can be expressed by the following form with $t$-independent coefficients $B$ and $C$
%
%
\begin{align}
\Delta_{\pm}(t, t_{\mathrm{sep}}; {\bm q})=
B e^{-\Delta E({\bm q},-{\bm q})t} \pm C e^{-\Delta E({\bm 0},{\bm q})(t_{\mathrm{sep}}-t)},
\end{align}
where the non-interaction estimates $\Delta E({\bm q},{\bm k})=E_{\pi}({\bm k})+E_{N}({\bm q}+{\bm k})-E_{N}({\bm q})$ 
may be used. 
Therefore, the time derivative of the $\pi N$ contribution $\Delta_{\pm}(t, t_{\mathrm{sep}}; {\bm q})$ may have 
the following property:
%
%
\begin{align}
\partial_4 \Delta_{\pm}(t, t_{\mathrm{sep}}; {\bm q})=-E_\pi({\bm q}) \Delta_{\mp}(t, t_{\mathrm{sep}}; {\bm q})+(E_N({\bm q})-M_N)\Delta_{\pm}(t, t_{\mathrm{sep}}; {\bm q}),
\end{align}
which offers us to separate the $\pi N$ contribution $\Delta_{\pm}(t, t_{\mathrm{sep}}; {\bm q})$ from the $\widetilde{F}_P$ form factor using the time-derivative of the ratio correlator $\partial_4 R_{A_\alpha}^{5z}(t; \bm{q})$. 
Hereafter the nucleon energy $E_N(\bm{q})$ and the pion energy $E_\pi(\bm{q})$ are simply abbreviated by shorthand notations $E_N$ and $E_\pi$, respectively. 

The new method for determining $\widetilde{F}_P(q^2)$, including the time derivative of 
the $A_4$ and $A_i$ correlators, is given by
%
%
\begin{align}
\label{eq:new_FP}
\widetilde{F}_P(q^2)
&=-K\frac{\tilde{\cal R}^{5z}_{A_i}(t, {\bm q})}{q_iq_3}
+\frac{K}{\Delta E_N^2-E_\pi^2}
\left[
\Delta E_N\frac{\partial_4 \tilde{\cal R}^{5z}_{A_i}(t, {\bm q})}{q_iq_3}
+\frac{\partial_4 {\cal R}^{5z}_{A_4}(t, {\bm q})}{iq_3}
\right] 
\end{align}
with $\Delta E_N\equiv E_N-M_N$ and $K=\sqrt{2E_N(E_N+M_N)}$.
The first term corresponds to $\widetilde{F}_P^{\mathrm{std}}(q^2)$ in the standard method. The leading $\pi N$ contributions represented in terms of $\Delta_{+}(t, t_{\mathrm{sep}}; {\bm q})$ 
and $\Delta_{-}(t, t_{\mathrm{sep}}; {\bm q})$ can be completely eliminated by adding the second term in Eq.~(\ref{eq:new_FP}). For the ground state contribution, namely $\widetilde{F}_P(q^2)$, 
Eq.~(\ref{eq:new_FP}) is just a harmless linear combination exploiting the redundancy in the determination of
 $\widetilde{F}_P(q^2)$ from both of $C_{A_4}^{5z}(t; \bm{q})$ and ${C}_{A_i}^{5z}(t; \bm{q})$.
Therefore, if Eq.~(\ref{eq:new_FP}) successfully shows good plateau behavior, independent of the choice of $t_{\mathrm{sep}}$, it guarantees that the ground state contribution can be read accurately without excited-state contamination.

Since the $\widetilde{G}_P(q^2)$ was also observed to be strongly contaminated from the excited state, similar to $\widetilde{F}_P(q^2)$, in our previous works~\cite{{Ishikawa:2018rew},{Shintani:2018ozy},{Ishikawa:2021eut},{Tsuji:2023llh}}, we simply assume that  
%
%
\begin{align}
{\cal R}^{5z}_{P}(t,{\bm q})
&=iq_3K^{-1}
\left[\widetilde{G}_P(q^2)-\Delta_{P}(t, t_{\mathrm{sep}}; {\bm q}) \right],
\end{align}
where $\Delta_{P}(t, t_{\mathrm{sep}}; {\bm q})$ encodes the leading $\pi N$ state contributions that
cause a residual $t$-dependence in ${\cal R}_P^{5z}(t; \bm{q})$.
Unlike in the case of the axial-vector currents, only a single correlator cannot remove the $\pi N$ contribution $\Delta_{P}(t, t_{\mathrm{sep}}; {\bm q})$.
Instead, it was found that the axial Ward-Takahashi identity is well satisfied in terms of the 3pt functions of the nucleon in our previous study~\cite{Tsuji:2023llh}:
%
%
\begin{align}
\label{eq:AWTI}
Z_A [
\partial_\alpha C_{A_{\alpha}}^{5z}(t; \bm{q})
]=2 m_{\rm PCAC} C_{P}^{5z}(t; \bm{q}),
\end{align}
where $m_{\rm PCAC}$ corresponds to the bare quark mass which coincides with
the value determined from the pion 2pt functions~\cite{Tsuji:2023llh}. 
Recall that Eq.~(\ref{eq:AWTI}) is satisfied without isolating the ground-state
contribution from the excited-state contributions~\cite{Tsuji:2023llh}. 
Thus, Eq.~(\ref{eq:AWTI}) leads to the following PCAC relation for the leading $\pi N$ contributions involved in $\tilde{\cal R}_{A_i}^{5z}(t; \bm{q})$ and 
${\cal R}_P^{5z}(t; \bm{q})$:
%
%
\begin{align}
\label{eq:piN_AWTI}
\Delta_{P}(t, t_{\mathrm{sep}}; {\bm q})=Z_A \frac{M_\pi^2}{2m_{\rm PCAC}}\Delta_{+}(t, t_{\mathrm{sep}}; {\bm q}),
\end{align}
which offers a simple subtraction method for determining the $\widetilde{G}_P(q^2)$ as below
%
%
\begin{align}
\label{eq:new_GP}
\widetilde{G}_P(q^2)
&=K\frac{{\cal R}^{5z}_{P}(t, {\bm q})}{iq_3}
+\frac{Z_A B_0 K}{\Delta E_N^2-E_\pi^2}
\left[
\Delta E_N\frac{
\partial_4 \tilde{\cal R}^{5z}_{A_i}(t, {\bm q})
}{q_iq_3}
+\frac{\partial_4 {\cal R}^{5z}_{A_4}(t, {\bm q})}{iq_3}
\right] 
\end{align}
with $B_0=\frac{M_\pi^2}{2m_{\mathrm{PCAC}}}$.
The first term corresponds to $\widetilde{G}_P^{\mathrm{std}}(q^2)$ in the standard method.

%
%
\begin{table}[b]
\caption{
Summary of simulation parameters in 2+1 flavor PACS10 ensembles with two different lattice spacings.
See Refs.~\cite{{Ishikawa:2018jee},{PACS:2019ofv},{Tsuji:2023llh}} for further details.
\label{tab:simulation_details}}
\centering
\begin{tabular}{cccccccccc}
\hline \hline
 $\beta$ & $L^3\times T$ &$\kappa_{ud}$ & $\kappa_{s}$ & $c_{\mathrm{SW}}$ & $a^{-1}$ [GeV] 
 & $M_\pi$ [GeV]
 & $Z_A^{\mathrm{SF}}$ \\
\hline
1.82& $128^3\times 128$ & 0.126117 & 0.124902 & 1.11 & 2.3 
&  135
& 0.9650(68) \\
2.00& $160^3\times 160$ & 0.125814& 0.124925 & 1.02 & 3.1
&  138
& 0.9783(21)\\
\hline \hline
\end{tabular}
\end{table}
%

\section{Numerical results}
In this study, we reanalyze the data sets generated in Refs.~\cite{{Shintani:2018ozy},{Tsuji:2023llh}} for $\widetilde{F}_{P}(q^2)$ and  $\widetilde{G}_{P}(q^2)$ using the new method described in Sec.~\ref{sec:New_method}.
The two data sets are computed with the first and second PACS10 ensembles, which are two sets
of gauge configurations generated in a large volume of about $(10\;\mathrm{fm})^4$ 
by the PACS Collaboration with the six stout-smeared  
${\cal O}(a)$ improved Wilson-clover quark action and Iwasaki gauge action at $\beta=1.82$ and 2.00 corresponding
to the lattice spacings of 0.09 fm (coarse) and 0.06 fm (fine), respectively~\cite{{Shintani:2018ozy},{Tsuji:2023llh}}. 
A brief summary of the simulation parameters is given in Table~\ref{tab:simulation_details}.
The simulated pion masses on both lattices are almost at the physical point.

%
%
\begin{figure*}[ht]
\centering
\includegraphics[width=0.49\textwidth,bb=0 0 864 720,clip]
{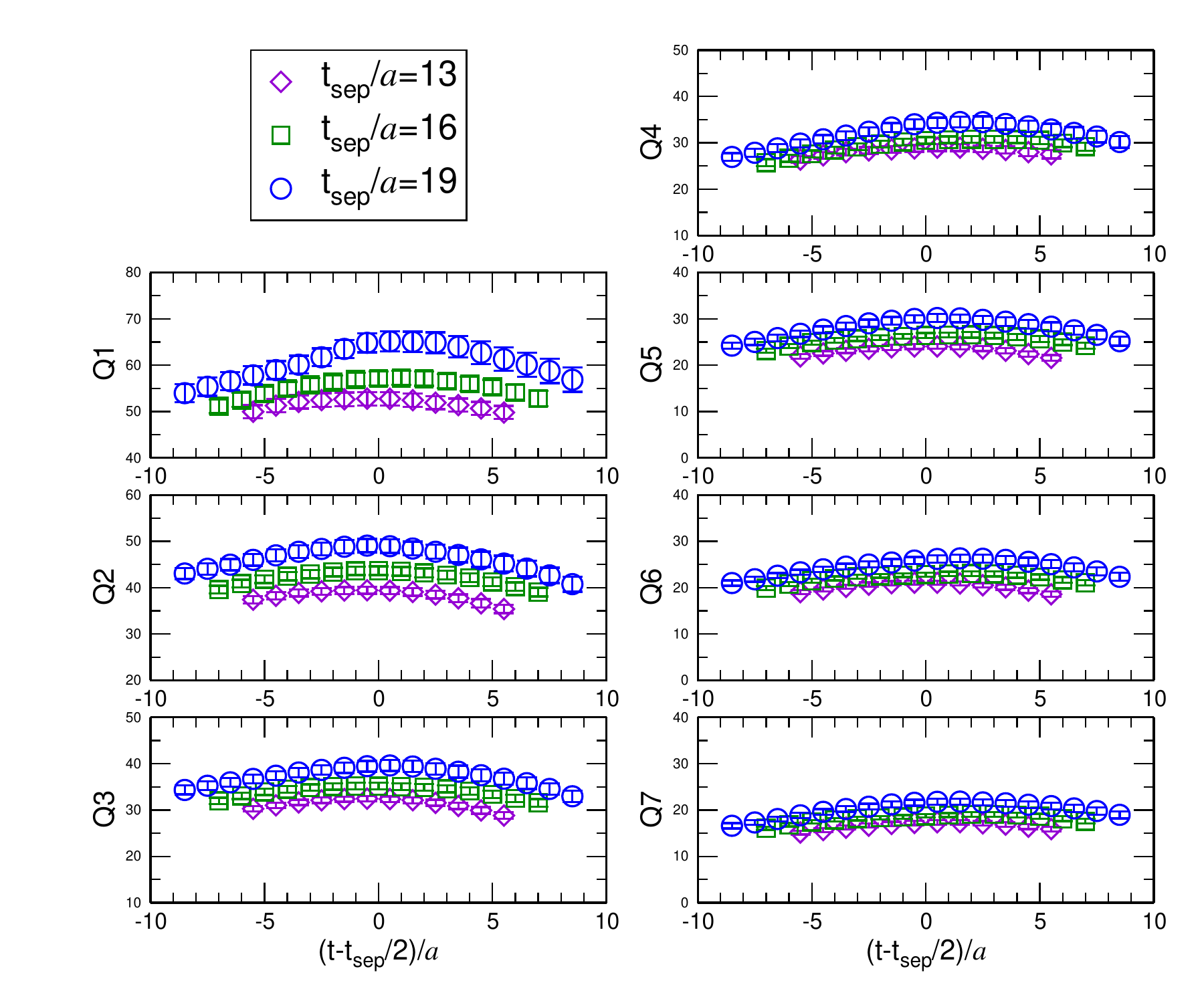}
\includegraphics[width=0.49\textwidth,bb=0 0 864 720,clip]
{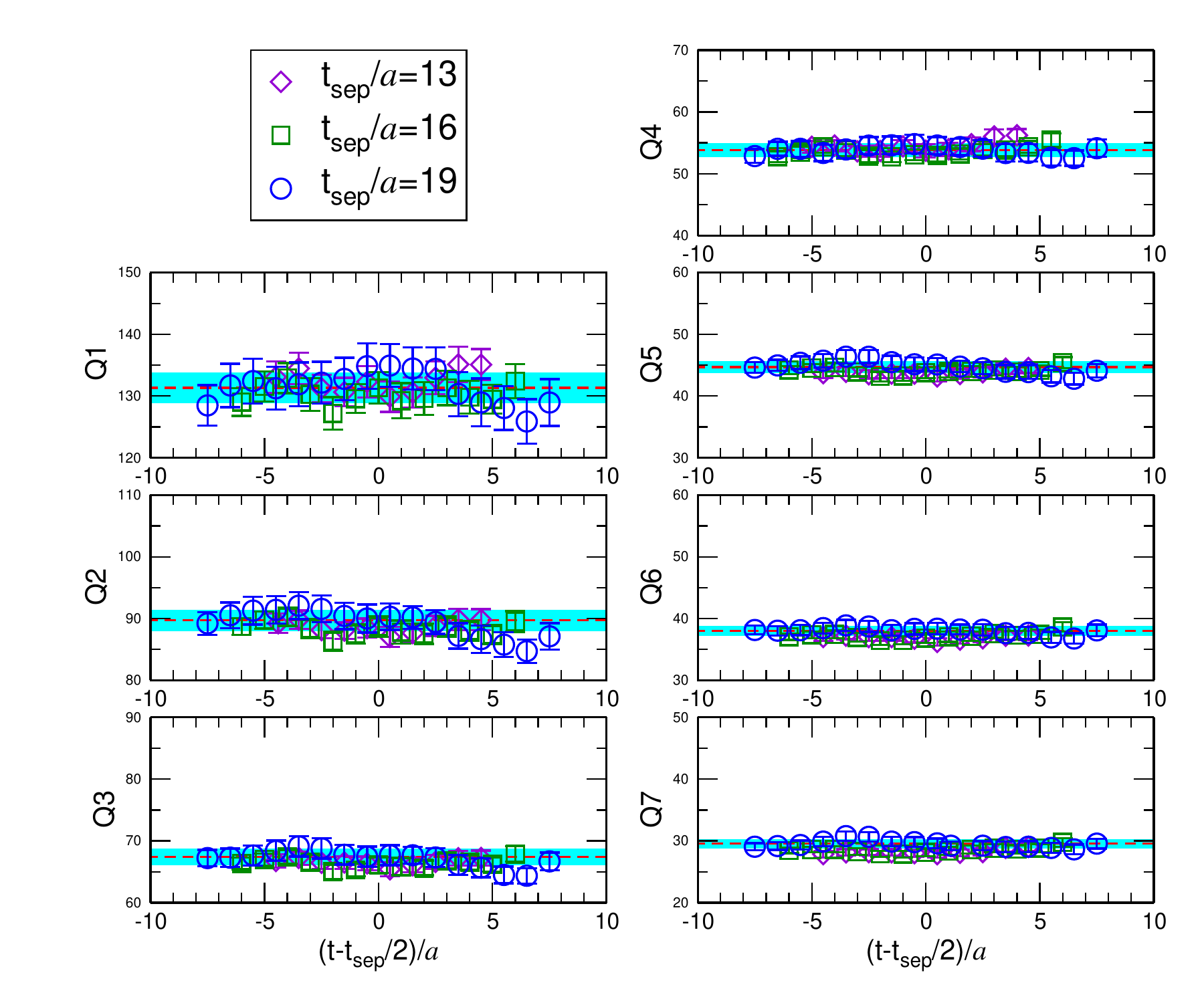}
\caption{
The values of $2M_N F_P^{\rm std}$ (left) and $2M_N F_P$ (right)
computed using the second PACS10 ensemble ($160^4$ lattice)
with $t_{\mathrm{sep}}/a=13$ (diamonds), $16$ (squares) and $19$ (circles) 
for all momentum transfers
as functions of the current insertion time slice $t$. In the right panel, the horizontal bands are calculated from the PPD model ($2M_N F_P^{\rm PPD}(q^2)$).
}
\label{fig:F_P_plateau_comp}
\end{figure*}

%
%
%
\begin{figure*}[ht]
\centering
\includegraphics[width=0.49\textwidth,bb=0 0 864 720,clip]
{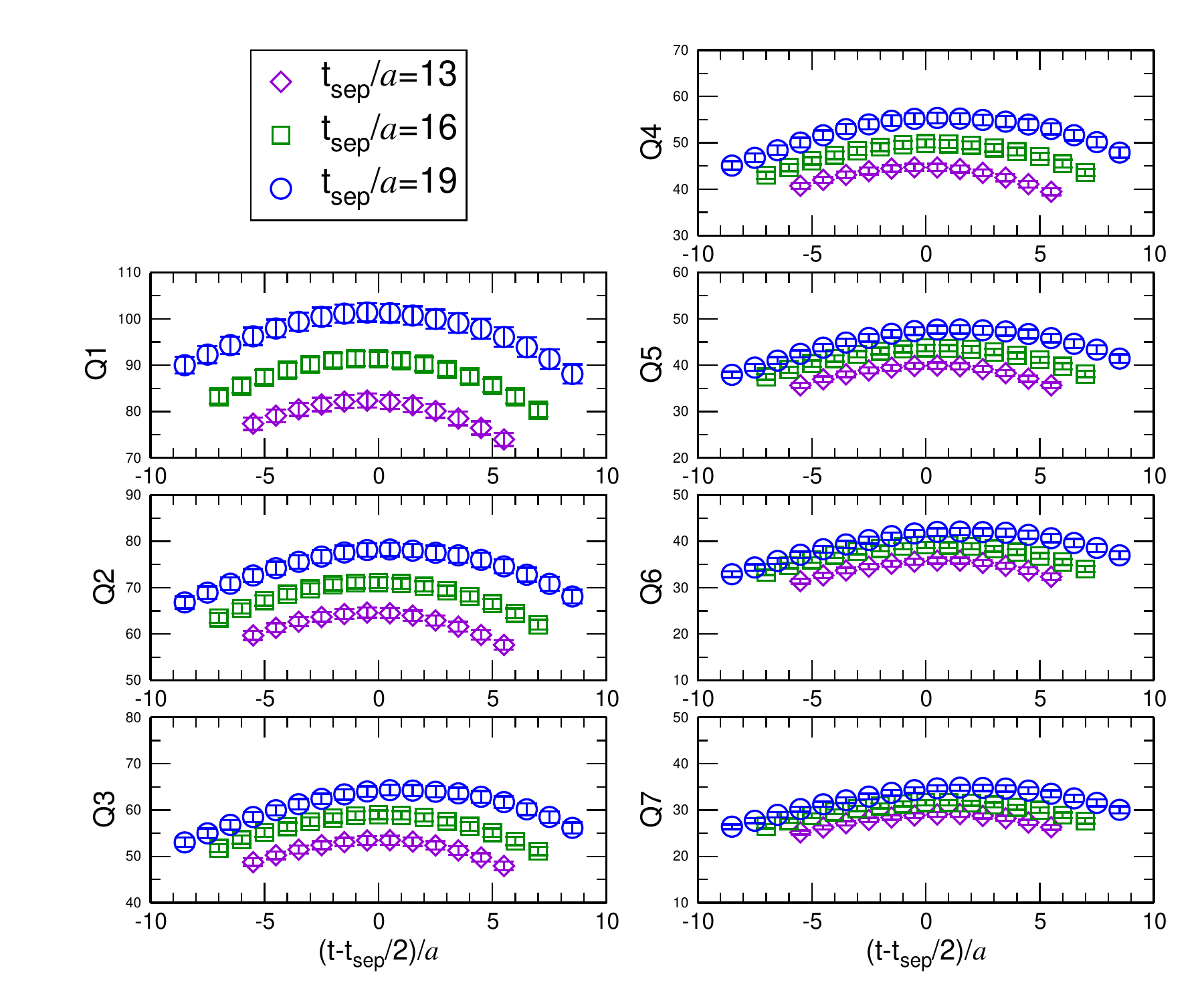}
\includegraphics[width=0.49\textwidth,bb=0 0 864 720,clip]
{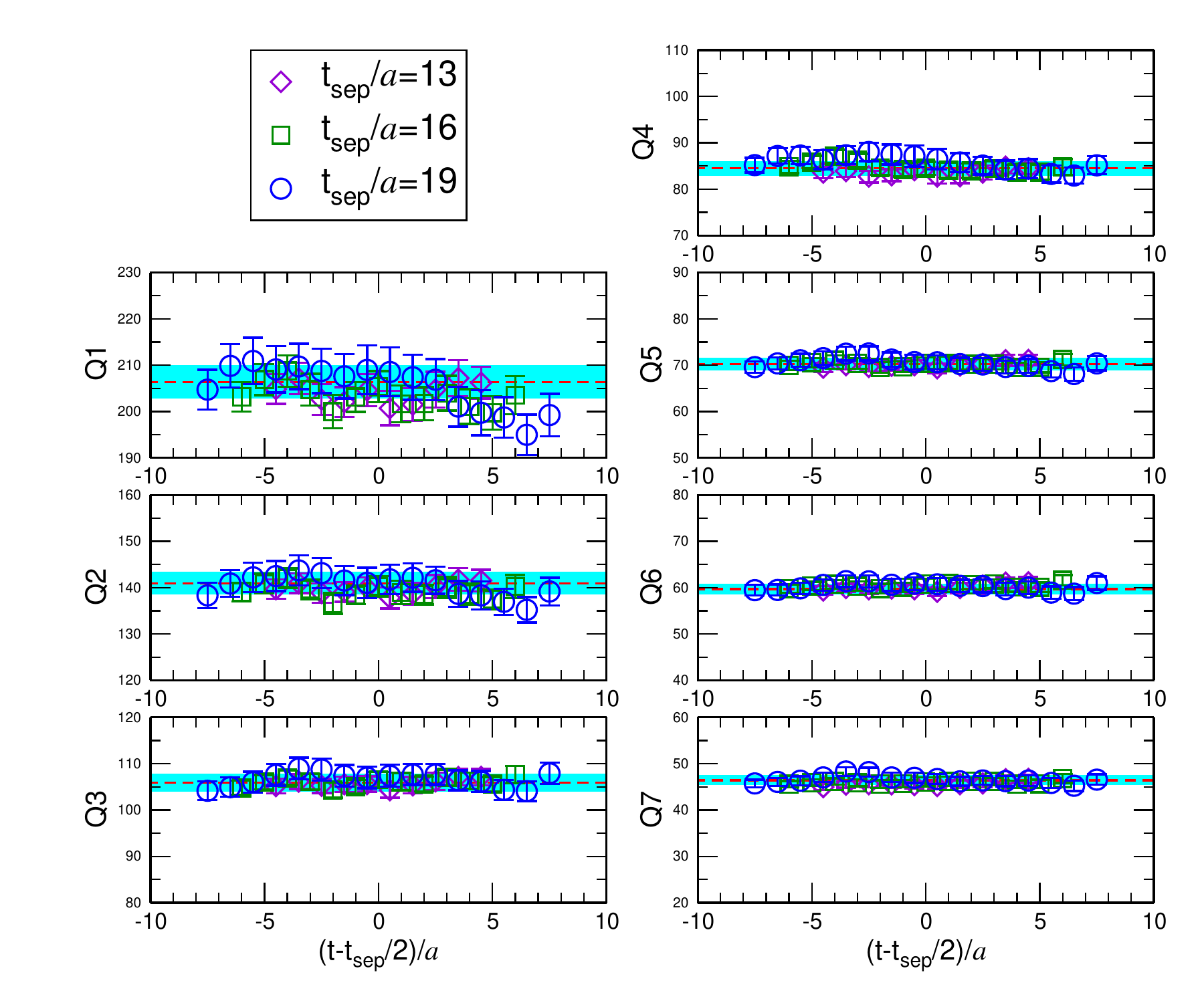}
\caption{
The values of $\widetilde{G}_P^{\rm std}$ (left) and $\widetilde{G}_P$ (right)
computed with $t_{\mathrm{sep}}/a=13$ (diamonds), $16$ (squares) and $19$ (circles) for all momentum transfers as functions of the current insertion time slice $t$. In the right panel, the horizontal bands are calculated from the PPD model ($\widetilde{G}_P^{\rm PPD}(q^2)$).
}
\label{fig:G_P_plateau_comp}
\end{figure*}

The $F_P$ form factor is extracted from Eq.~(\ref{eq:new_FP}) as a function of the current insertion time $t$.
In Fig.~\ref{fig:F_P_plateau_comp}, we compare the $t$-dependence
and $t_{\mathrm{sep}}$-dependence of $F_P(q^2)$ obtained by both the standard (left panel) and the simple subtraction (right panel) methods for the $160^4$ lattice ensemble. The new method is really effective in obtaining an asymptotic plateau in all cases of $t_{\mathrm{sep}}/a=\{13,16,19\}$ for all $q^2$. 
Indeed, as shown in the right panel, the $t$-dependence is eliminated and the $t_{\mathrm{sep}}$-dependence is not visible either. Furthermore, the plateau values are
consistent with the PPD model.

\begin{figure*}[h]
\centering
\includegraphics[width=0.49\textwidth,bb=0 0 792 612,clip]{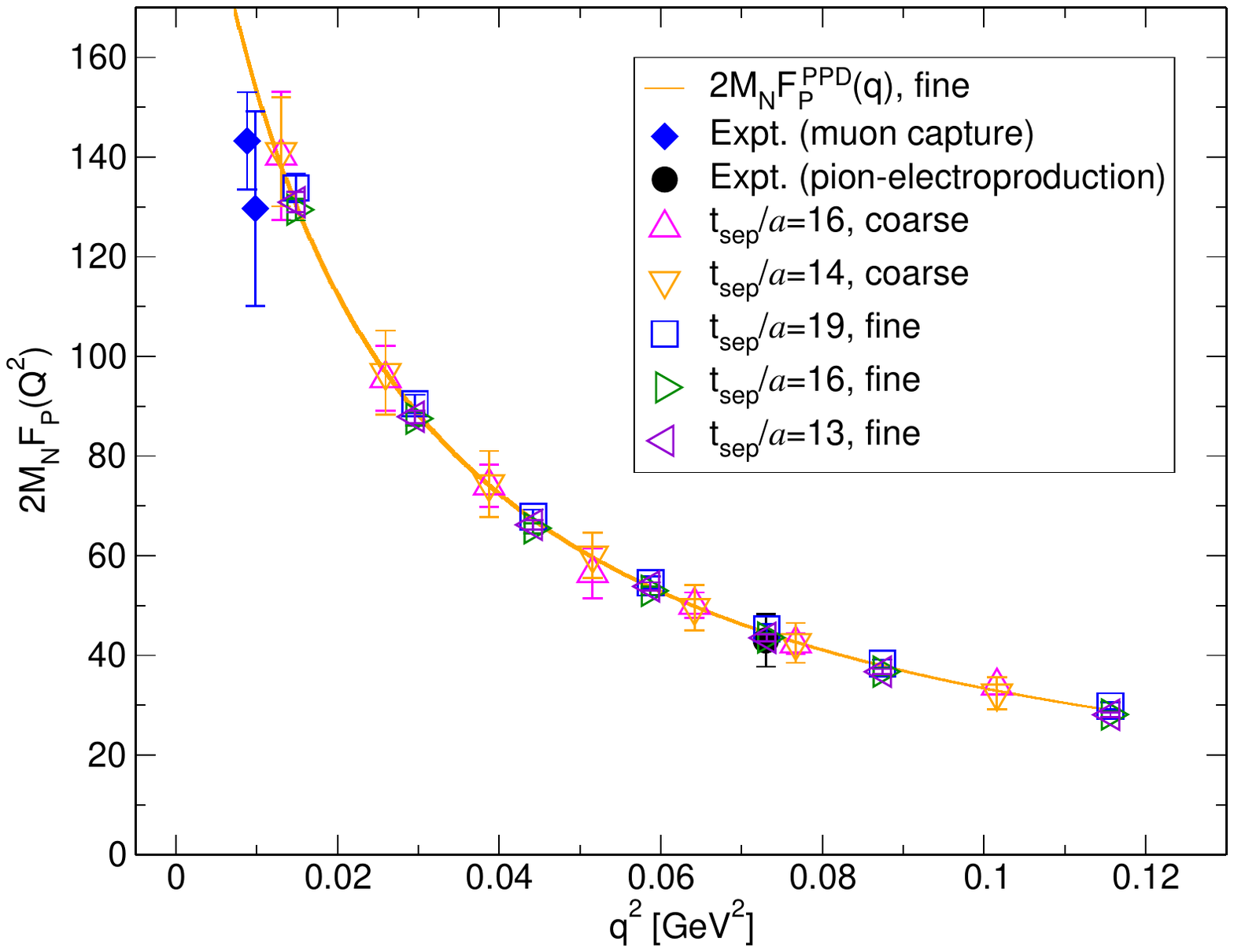}
\includegraphics[width=0.49\textwidth,bb=0 0 792 612,clip]{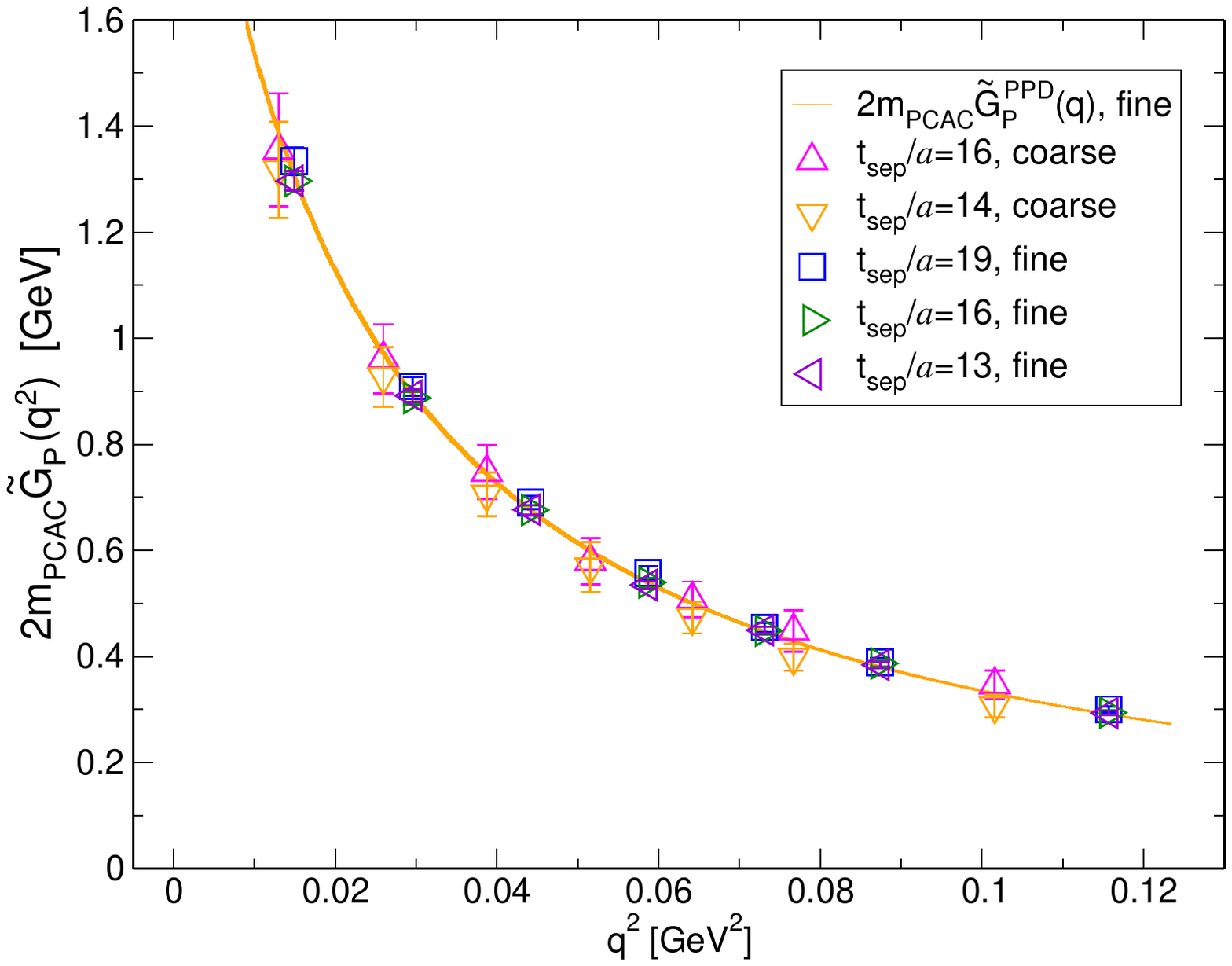}
\caption{Results of $2M_N F_P(q^2)$ (left panel) and $2m_{\mathrm{PCAC}} \widetilde{G}_P(q^2)$ (right panel) obtained by the new method as a function of $q^2$. In each panel, the solid curve is given by
the PPD model defined in Eq.~(\ref{eq:PPD}).}
\label{fig:tsep_dep_fp_gp}
\end{figure*}

As shown in Fig.\ref{fig:G_P_plateau_comp}, similarly for the $G_P$ form factor, the new method eliminates the slight convex shape associated with the excited-state contribution and yields a plateau behavior consistent with the PPD model without $t_{\mathrm{sep}}$ dependence as well. In Fig.~\ref{fig:tsep_dep_fp_gp}, we plot the $q^2$ dependence of
$2 M_N F_P(q^2)$ (left panel) and $2 m_{\mathrm{PCAC}} \widetilde{G}_P(q^2)$ (right panel)
for all data sets of $160^4$ and $128^4$ lattices together with experimental data points from muon capture~\cite{Gorringe:2002xx} and pion-electro production~\cite{Choi:1993vt}.  

%
%

\begin{figure*}[h]
\centering
\includegraphics[width=0.49\textwidth,bb=0 0 792 612,clip]{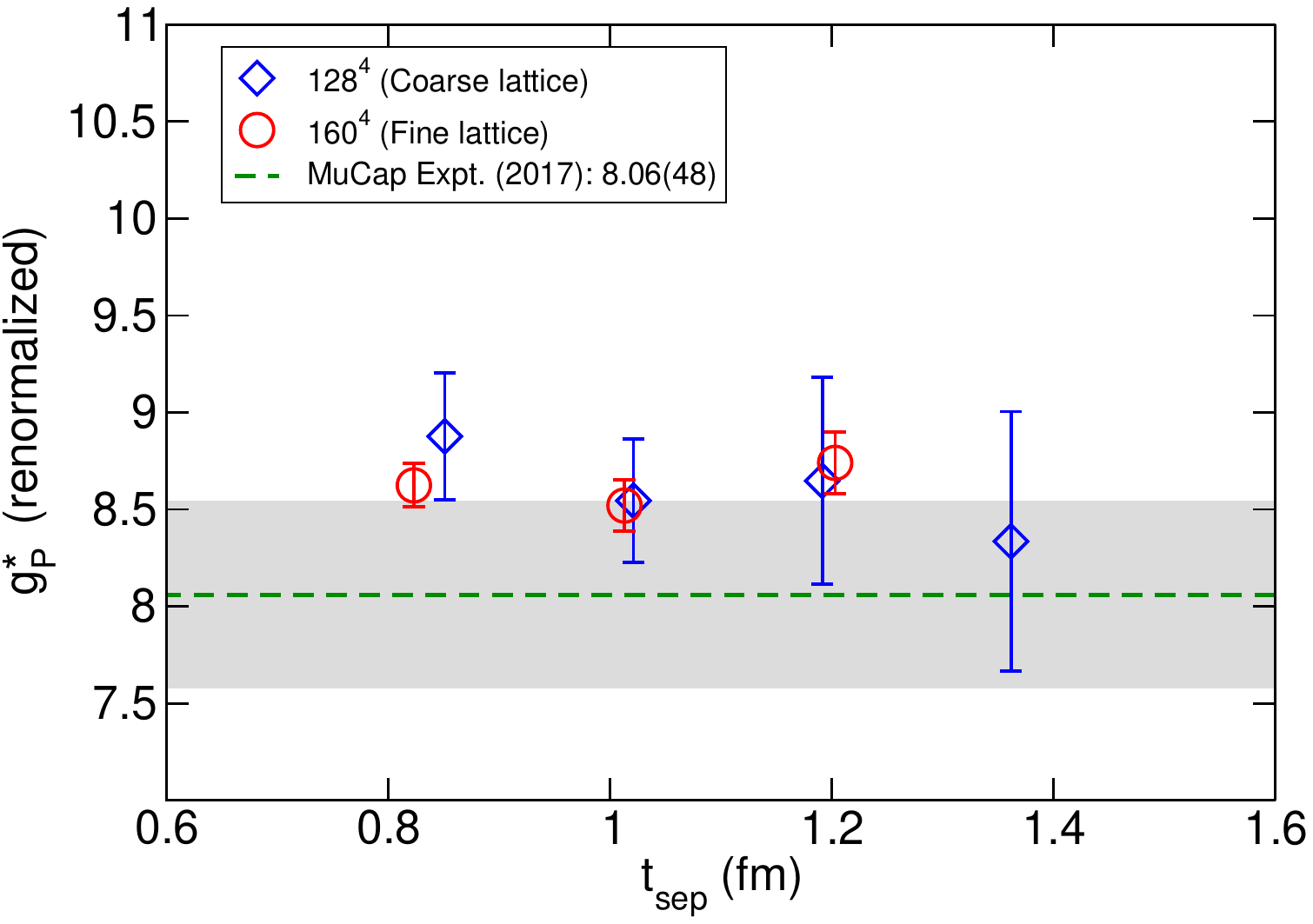}
\includegraphics[width=0.49\textwidth,bb=0 0 792 612,clip]{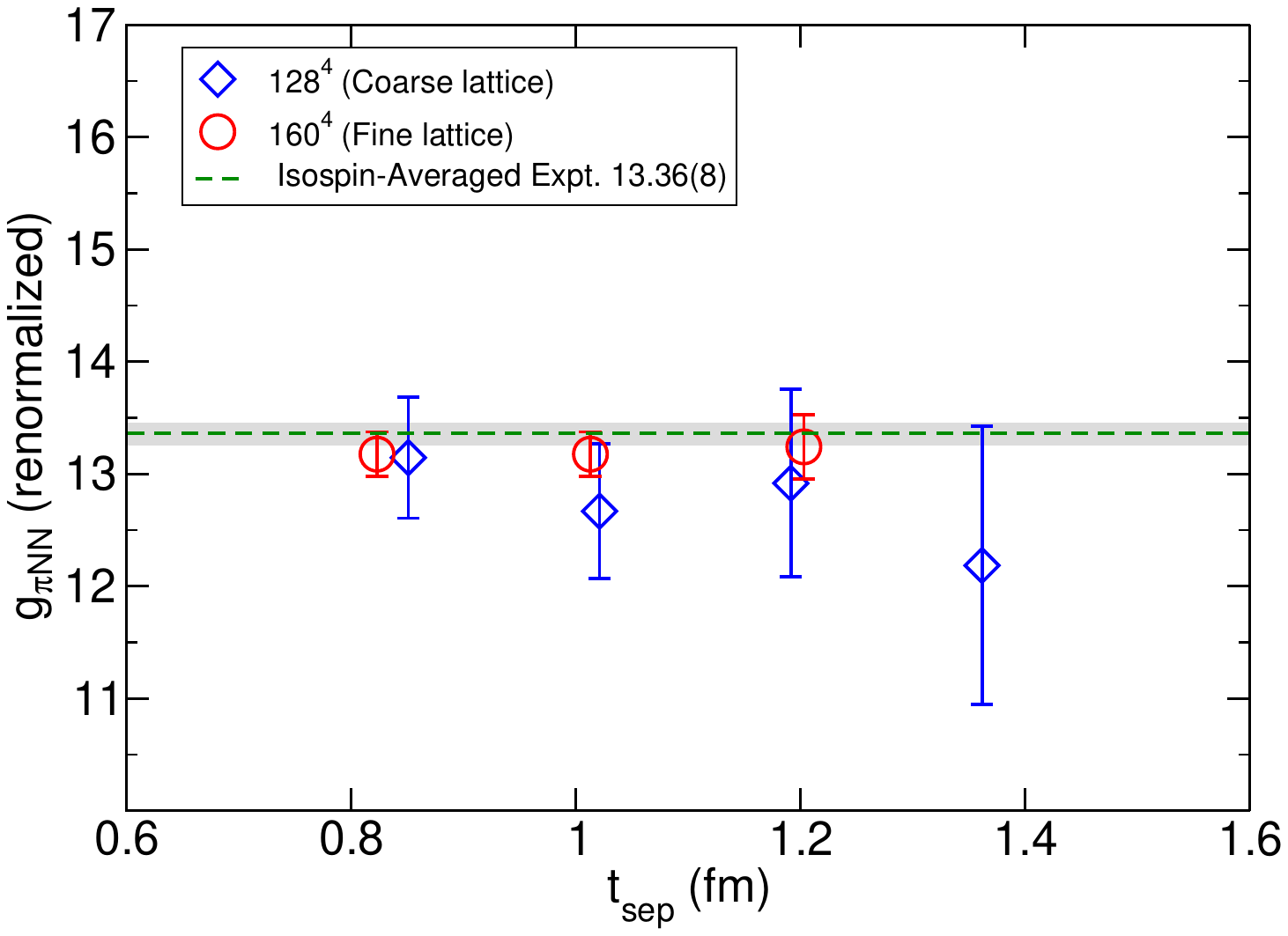}
\caption{
The source-sink separation ($t_{\rm sep}$) dependence of the renormalized
values of $g_P^{\ast}$ (left) and $g_{\pi NN}$ (right). In each panel, the horizontal axis gives $t_{\rm sep}$ in physical units, while the horizontal dashed line with the gray band denotes the experimental value. 
}
\label{fig:tsep_dep_gp_gpNN}
\end{figure*}

Next, $g_P^\ast$ and $g_{\pi NN}$ are evaluated from the obtained $F_P$ form factor according to the $q^2$ dependence analysis based on the z-expansion method applied to $(q^2+m_\pi^2)F_P(q^2)$. As shown in Fig~\ref{fig:tsep_dep_gp_gpNN}, no $t_{\mathrm{sep}}$ dependences are visible for either case
and the discretization error on these quantities is less than
3-4 \%, which is well controlled in our calculations as well as $g_A$. More importantly, the evaluation is possible with much smaller errors than the experimental values for $g_P^\ast$ and comparable errors for $g_{\pi NN}$. 
Summary plots of our results together with the experimental values and other lattice QCD results for $g_P^\ast$ and $g_{\pi NN}$ can be found in Ref.~\cite{Tsuji:2024scy}. 

%
%

\section{Summary}
We have studied nucleon form factors in the axial-vector and pseudo-scalar channels in 2+1 flavor QCD using two sets of the PACS10 configurations at coarse and fine lattice spacings. Our simulations were carried out in very large spatial volumes, which allow us to access the low $q^2$ region, at the physical point essential for low-energy chiral behavior. The nucleon interpolating operator has been adopted with well-tuned smearing parameters that guarantee ground-state dominance in $F_A(q^2)$, although the two types of pseudo-scalar form factors, $F_P(q^2)$ and $G_P(q^2)$, still suffer from the excited-state contamination. 
In this study, we thus propose a simple subtraction method for removing the so-called leading $\pi N$-state contamination induced by the pion-pole structure appears in $F_P(q^2)$ and $G_P(q^2)$.
The new method achieves the following points: 1) it can use the 3pt-functions of both spatial and temporal axial-vector currents to determine $F_P(q^2)$, 2) it is applicable for $G_P(q^2)$ with a help of the PCAC relation, 3) 
it eliminates both $t$-dependence and $t_{\mathrm{sep}}$-dependence in both $F_P(q^2)$ and $G_P(q^2)$, 4) it makes results compatible with both the experiment and the PPD model, and 5) it provides more accurate results of two target quantities, $g_P^\ast$ and $g_{\pi NN}$, comparing the multi-state analysis that was used in 
other groups~\cite{{RQCD:2019jai},{Park:2021ypf},{Jang:2023zts},{Alexandrou:2023qbg}}.

\acknowledgments
We would like to thank members of the PACS collaboration for useful discussions.
K.-I.~I. is supported in part by MEXT as ``Feasibility studies for the next-generation computing infrastructure".
Numerical calculations in this work were performed on Oakforest-PACS in Joint Center for Advanced High Performance Computing (JCAHPC) and Cygnus  and Pegasus in Center for Computational Sciences at University of Tsukuba under Multidisciplinary Cooperative Research Program of Center for Computational Sciences, University of Tsukuba, and Wisteria/BDEC-01 in the Information Technology Center, the University of Tokyo. 
This research also used computational resources of the K computer (Project ID: hp180126) and the Supercomputer Fugaku (Project ID: hp20018, hp210088, hp230007, hp230199, hp240028, hp240207) provided by RIKEN Center for Computational Science (R-CCS), as well as Oakforest-PACS (Project ID: hp170022, hp180051, hp180072, hp190025, hp190081, hp200062),  Wisteria/BDEC-01 Odyssey (Project ID: hp220050) provided by the Information Technology Center of the University of Tokyo / JCAHPC.
The  calculation employed OpenQCD system(http://luscher.web.cern.ch/luscher/openQCD/). 
This work is supported by the JLDG constructed over the SINET5 of NII,
This work was also supported in part by Grants-in-Aid for Scientific Research from the Ministry of Education, Culture, Sports, Science and Technology (Nos. 18K03605, 19H01892, 22K03612, 23H01195, 23K03428, 23K25891) and MEXT as ``Program for Promoting Researches on the Supercomputer Fugaku'' (Search for physics beyond the standard model using large-scale lattice QCD simulation and development of AI technology toward next-generation lattice QCD; Grant Number JPMXP1020230409).

\bibliographystyle{JHEP}
\bibliography{skeleton}

\end{document}